# Improving Waiting Time of Tasks Scheduled Under Preemptive Round Robin Using Changeable Time Quantum

Saad Zagloul Rida[*] Safwat Helmy Hamad [**] Samih Mohemmed Mostafa [***]

[*]Faculty of Science, Mathematics Department, South Valley University, Egypt, Quena.

[**] Faculty of Computer and Information Sciences, Ain Shams University, Egypt, Cairo.

[***] Faculty of Science, Computer Science Department, South Valley University, Egypt, Quena.

## ABSTRACT

Minimizing waiting time for tasks waiting in the queue for execution is one of the important scheduling criteria which took a wide area in scheduling preemptive tasks. In this paper we present Changeable Time Quantum (CTQ) approach combined with the round-robin algorithm, we try to adjust the time quantum according to the burst times of the tasks in the ready queue. There are two important benefits of using (CTQ) approach: minimizing the average waiting time of the tasks, consequently minimizing the average turnaround time, and keeping the number of context switches as low as possible, consequently minimizing the scheduling overhead. In this paper, we consider the scheduling problem for preemptive tasks, where the time costs of these tasks are known a priori. Our experimental results demonstrate that CTQ can provide much lower scheduling overhead and better scheduling criteria.

**General Terms:** Algorithm, round-robin, CPU length.

**Keywords:** Processor sharing, proportional sharing, residual time, survived tasks, cyclic queue.

## 1. INTRODUCTION

Modern operating systems (OS) nowadays have become more complex than ever before. They have evolved from a single task, single user architecture to a multitasking environment in which tasks run in a concurrent manner. Allocating CPU to a task requires careful attention to assure fairness and avoid task starvation for CPU. Scheduling decision try to minimize the following: average turnaround time, average waiting time for tasks and the number of context switches [3]. Scheduling algorithms are the mechanism by which a resource is allocated to a client. In our research we restrict the concept of a resource to CPU time and clients to tasks. A scheduling decision refers to the concept of selecting the next process for execution. The scheduler runs the first task in the queue for the specified time quantum, which is the maximum time interval the task is allowed to run before another scheduling decision is made. Note that the time quantum is typically expressed in time units of constant size. As a result, we refer to the units of time quanta as time units (tu) in this paper rather than an absolute time measure such as seconds. During each scheduling decision, a context switch occurs, meaning that the current process will stop its execution and put back to the ready queue and another process will be dispatched. We define scheduling overhead as the incurred overhead when making a scheduling decision. One of the oldest, simplest and most widely used scheduling algorithms is round-robin. RR algorithms are widely used in modern OSs like Linux, BSD and Windows. All use multi-level feedback queues with priorities and a RR scheduler over each task queue. In addition, RR algorithms have low scheduling overhead of $O(1)$, which means scheduling the next task takes a constant time [5, 4, 10]. The performance of the RR algorithm depends heavily on the size of the time quantum. At one extreme, if the time quantum is extremely large, the RR policy is the same as the FCFS policy. If the time quantum is extremely small, the RR approach is called processor sharing [3]. In this paper we try to minimize waiting time, turnaround time and number of the context switches by changing time quantum according to the burst times of the tasks in the queue. Our results show that CTQ scheduler can provide better scheduling criteria than these other schedulers that dealing with fixed time quantum. Furthermore, our results show that CTQ achieves this minimization with lower scheduling overhead.

This paper presents the design and implementation of CTQ in which we try to minimize the scheduling criteria by adjusting the time quantum of the tasks with respect to their burst times. Section 2 discusses background. Section 3 presents the proposed algorithm. Section 4 presents the changeable consideration. Section 5 presents the simulation studies. Finally we present some concluding remarks and directions for future work.

## 2. BACKGROUND

The round-robin (RR) scheduling algorithm is designed especially for time-sharing systems. As known that the concept of processor sharing introduces a heuristics to prevent long jobs from blocking short jobs in queue. In practice round-robin has to be used. RR is also one of the oldest, simplest and most widely used proportional share scheduling algorithms, and because of its usefulness, many proportional share scheduling mechanisms have been developed [7, 1, 12, 6, 8, 2, 11, 9]. A modified version of RR is the Weighted Round Robin (WRR) in which each task T has a specified weight that specifies its share of the CPU time. If a time quantum $TQ$ is specified to be 10 time units (tu), and we have three tasks A, B, and C having weights 7, 4, and 9, then the time quantum given to each task is proportional to the task weight. An example of one round of WRR will assign, for example, task A 70% of the time quantum. Similarly task B will receive 40% of the time quantum and task C will receive 90% [13]. WRR is the simplest proportional-share scheduling algorithm. Weighted round-robin (WRR) provides proportional sharing by running all clients with the same frequency but adjusting the size of their time quanta. A more recent variant called deficit round-robin [11] has been developed for network packet scheduling with similar behavior to a weighted round-robin CPU scheduler. WRR is simple to implement and schedules clients in $O(1)$ time. Burst Round Robin (BRR) [13] is another weighting technique for round-robin CPU scheduler as an attempt to combine the low scheduling overhead of round-robin algorithms and favor short jobs.

## 3. THE PROPOSED ALGORITHM

In this section we introduce the equations that determine the time quantum $TQ$ that gives the smallest average waiting time in each round. $TQ$ is confined in the range from 1 to LBT, where LBT is the largest burst time of the tasks.

Equation 1:

$$NTQ(T_i) = \begin{cases} \left\lfloor \dfrac{BT_i}{TQ} \right\rfloor^{1} & if \quad BT_i \neq l * TQ \\ & l = 1,2,3,... \\ \dfrac{BT_i}{TQ} - 1 & if \quad BT_i = l * TQ \\ & l = 1,2,3,... \end{cases}$$

where $BT_i$ is the burst time of the task $T_i$, and $NTQ(T_i)$ is the number of times the task $T_i$ exploits the time quantum $TQ$. In the following example:

| TASK | BURST TIME |
|---|---|
| T1 | 24 |
| T2 | 3 |
| T3 | 3 |

if we use a time quantum of 4 time units, then we see from the Gantt Chart:

| T1 | T2 | T3 | T1 | T1 | T1 | T1 | T1 |
|---|---|---|---|---|---|---|---|

0 4 7 10 14 18 22 26 30

that the $NTQ(T_1)$ is 5, the $NTQ(T_2)$ is 0, and the $NTQ(T_3)$ is 0. But the number of context switches of $T_1$ is 1, the number of context switches of $T_2$ is 0, and the number of context switches of $T_3$ is 0.

Equation 2:

$$SLTQ(T_i) = \begin{cases} 0 + \sum_{k=1}^{i-1} \begin{cases} TQ \; if \; NTQ(T_k) > 0 \\ BT_k \; if \; NTQ(T_k) = 0 \end{cases} if \; NTQ(T_i) = 0 \\ NTQ(T_i) * TQ + \sum_{k=1, k \neq i}^{n} \begin{cases} BT_k \quad if \quad NTQ(T_k) < NTQ(T_i) \quad and \quad k \neq i \\ BT_k \quad if \quad NTQ(T_k) = NTQ(T_i) \quad and \quad k < i \\ (NTQ(T_i) * TQ) \;\; if \; NTQ(T_k) \geq NTQ(T_i) \, and \;\; k > i \\ (NTQ(T_i) + 1) * TQ \, if \;\; NTQ(T_k) > NTQ(T_i) \;\, and \, k < i \end{cases} if \; NTQ(T_k) > 0 \end{cases}$$

where $n$ is the number of the tasks, and $SLTQ(T_i)$ is the starting of the last time quantum of $T_i$. In the above example the $SLTQ(T_1)$ is 26, the $SLTQ(T_2)$ is 4, and the $SLTQ(T_3)$ is 7.

Equation 3:

$$WT(T_i) = SLTQ(T_i) - NTQ(T_i) * TQ$$

$$TWT = \sum_{i=1}^{n} WT(T_i)$$

$$AVGWT = TWT / n$$

where $WT(T_i)$ is the waiting time of task $T_i$, $TWT$ is the total waiting time of all tasks, and $AVGWT$ is the average waiting time of the tasks in the run queue.

## 4. THE CHANGEABLE CONSIDERATION

CTQ combines the benefit of low overhead round-robin scheduling with low average waiting time, this depends on the size of the preselected time quantum. The CTQ scheduling algorithm can be briefly described in the following steps:

1. Order the tasks in the run queue as FIFO manner.

2. Starting from the beginning of the run queue, run each task for one time quantum in a round-robin manner, until the task number n reaches. This is the first round in the cyclic queue.

3. Calculate the burst times for the survived tasks (*residual times*) in the next round, and implement the equations that determine the candidate time quantum.

4. Repeat the step 3 till there are no tasks waiting in the run queue.

Changing the time quantum in each round in the cyclic queue will give better results in the above criteria. Using the Changeable Time Quantum (CTQ) technique, in each round a different $TQ$ is used. If we have $n$ tasks in a round $r1$ and $m$ tasks that have burst times equal to or less than the time quantum used in $r1$, then there are $n-m$ tasks in the next round, where $n \geq m$.
The residual time of the task $T_i$ in the round number $q$ is determined from the equation:

$$\text{Re}\,sidual_Time[T_i] = Burst_Time(T_i) - \sum_{k=1}^{q-1} TQ[k]$$

where $TQ[k]$ is the time quantum in the round number $k$. In each successive round we implement the equations with respect to the residual times of the survived tasks. To demonstrate the previous consideration, we take the following example:

Consider the following set of tasks that arrive at time 0, each of which with the length of the CPU burst time.

| TASK | BURST TIME |
|---|---|
| T1 | 20 |
| T2 | 20 |
| T3 | 5 |
| T4 | 3 |
| T5 | 1 |

If we use the time quantum equal to 1 tu in the round-robin, this gives:

THE AVERAGE WAITING TIME = 17

THE AVERAGE TURNAROUND TIME = 26.8

THE TOTAL NUMBER OF CONTEXT SWITCHES IS 44

When we apply the (CTQ) technique: if the time quantum in the first round is equal to 1 ($TQ[1] = 1$)

(ROUND NO. 1)
($TQ[1] = 1$)

| T1 | T2 | T3 | T4 | T5 |
|---|---|---|---|---|
| 0–1 | 1–2 | 2–3 | 3–4 | 4–5 |

0 1 2 3 4 5

The survived tasks are T1, T2, T3, and T4 each of which with the length of the CPU burst time:

| TASK | RESIDUAL TIME |
|---|---|
| T1 | 19 |
| T2 | 19 |
| T3 | 4 |
| T4 | 2 |
| | |

After implementing the equations, we obtain:

```
TQ = 2
```

After using $TQ[2]$ equal to 2, the Gantt Chart is:

(ROUND NO. 2)
($TQ[2]$ =2)

| T1 | T2 | T3 | T4 |
|---|---|---|---|

5 7 9 11 13

from the survived tasks:

| TASK | RESIDUAL TIME |
|---|---|
| T1 | 17 |
| T2 | 17 |
| T3 | 2 |
| | |
| | |

the equations give:

```
TQ = 2
```

Again we use $TQ[3]$ equal to 2, the Gantt Chart is:

(ROUND NO. 3)
($TQ[3]$ =2)

| T1 | T2 | T3 |
|---|---|---|

13 15 17 19

from the survived tasks:

| TASK | RESIDUAL TIME |
|---|---|
| T1 | 15 |
| T2 | 15 |
| | |
| | |
| | |

in the last round we have $TQ[4]$ =15, the Gantt Chart is:

(ROUND NO. 4)
(TQ [4] = 15)

| T1 | T2 |
|---|---|

19 34 49

THE AVERAGE WAITING TIME = (14 + 29 + 14 + 10 + 4) / 5 = 14.2

THE AVERAGE TURNAROUND TIME = (34 + 49 + 19 + 13 + 5) / 5 = 24

THE TOTAL NUMBER OF CONTEXT SWITCHES IS (3 + 3 + 2 + 1 +0) = 9.

## 5. SIMULATION STUDIES

To demonstrate the effectiveness of the CTQ, we built a scheduling simulator that is a user-space program which takes four inputs, the scheduling algorithm, the number of tasks, the burst time of each task, and the first time quantum that will be used in the traditional round-robin, this time quantum will be selected to give the smallest average waiting time in the fixed round-robin. The simulator randomly assigns burst times to tasks.

To measure the effectiveness, we ran simulations for the proposed algorithm against fixed round-robin algorithm considered on 30 different combinations of $n$ and $BT's$, the burst times of the tasks varying from 1 to 500 tu. For each set of $(n, BT)$, we ran different number of tasks with different CPU lengths. Our results show a significant improvement in waiting time, turnaround time, and context switches as shown in figures 1, 2, and 3.

We also ran simulations for the proposed algorithm against the BRR [13] algorithm and fixed round-robin considered on 30 different combinations of $n$ and $BT's$, the burst times of the tasks varying from 1 to 100 tu, and the time quantum is 10 tu as proposed in BRR technique. For each set of $(n, BT)$, we ran different number of tasks with different CPU lengths. Figures 4, 5, and 6 show the comparative result between CTQ, BRR, and fixed RR.

The data produced by our simulations show that CTQ has the smallest average waiting time, average turnaround time, and number of context switches.

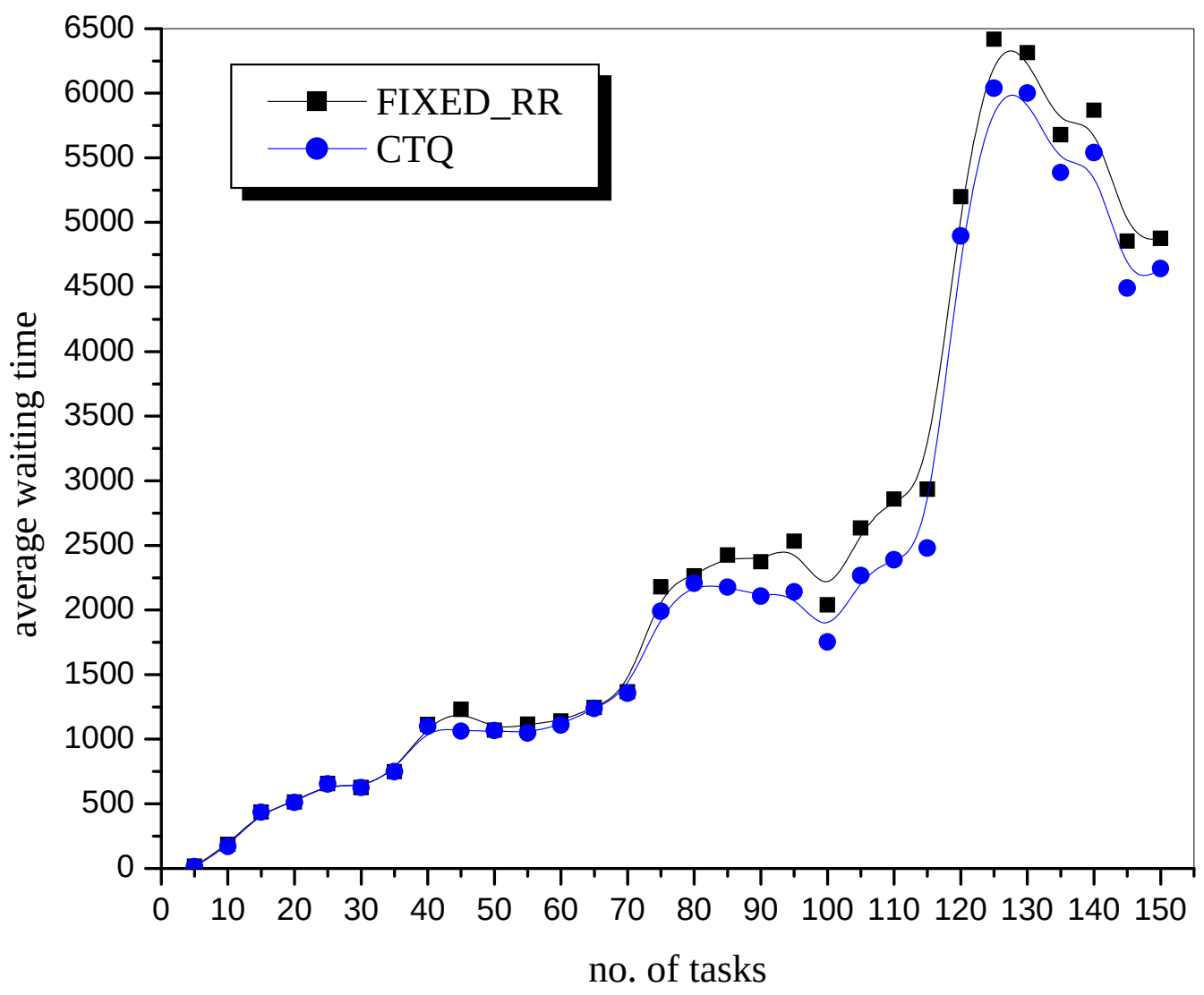

**Figures 1: Average waiting time for Fixed_RR and CTQ.**

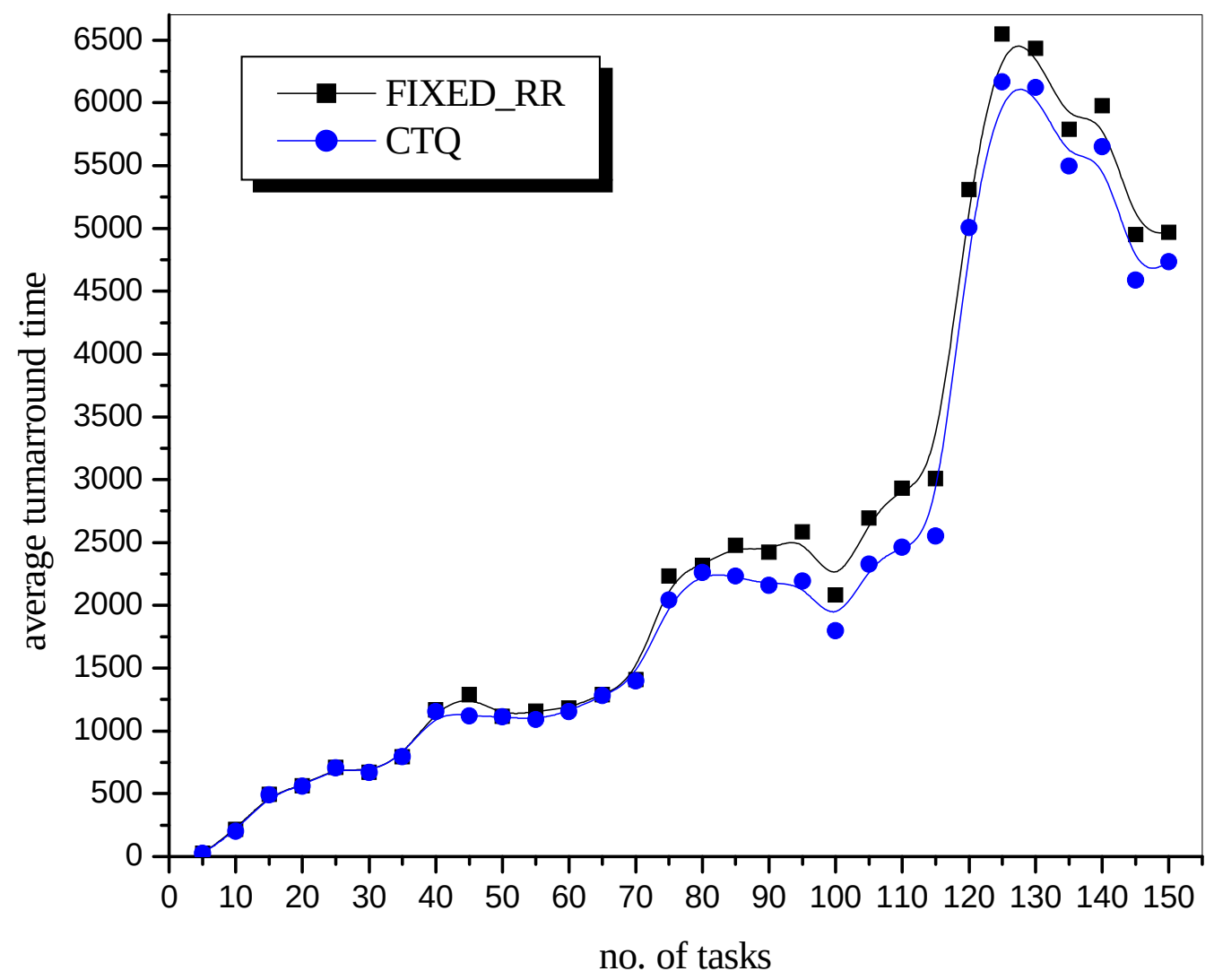

**Figures 2: Average turnaround time for Fixed_RR and CTQ.**

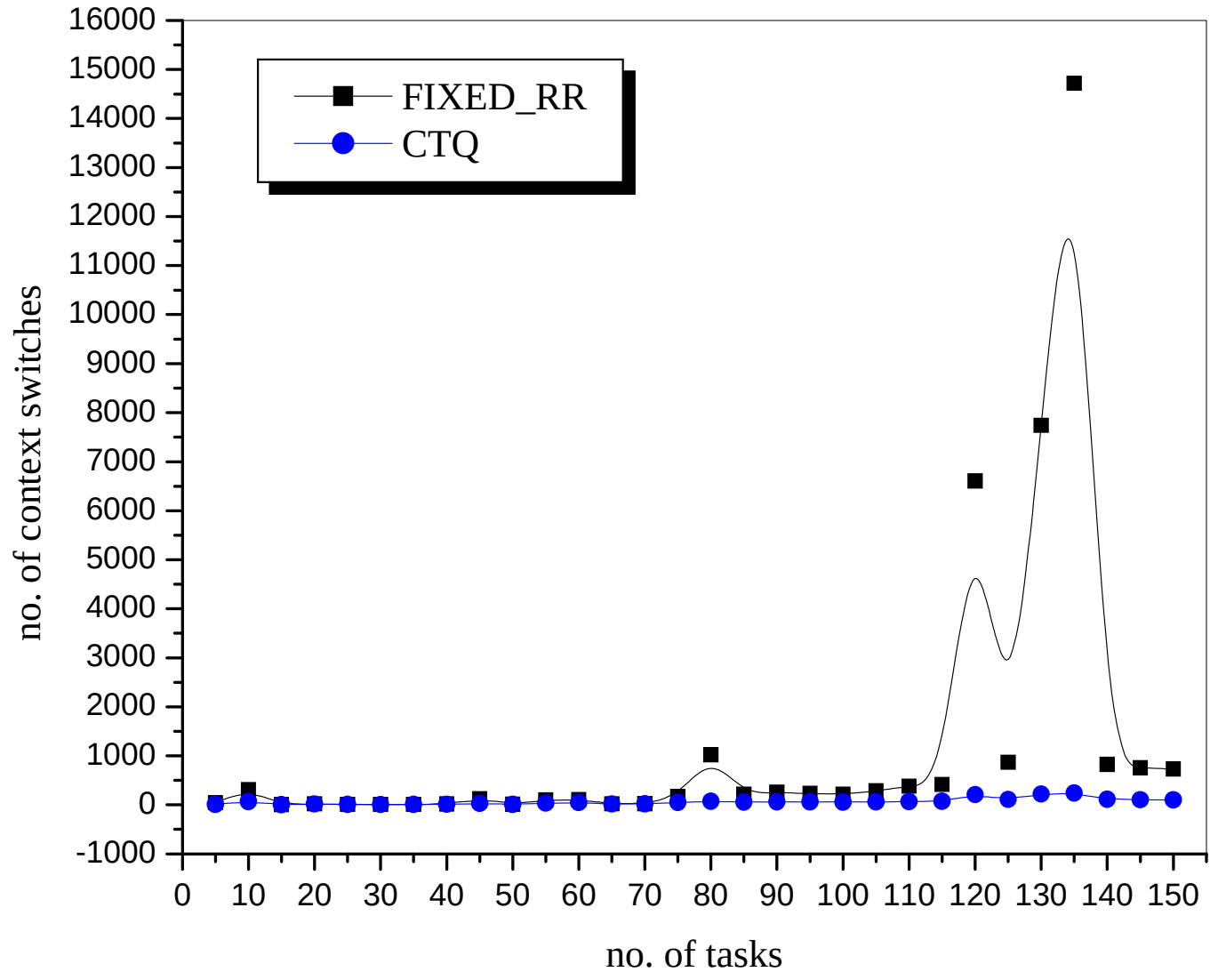

**Figures 3: Number of context switche for Fixed_RR and CTQ.**

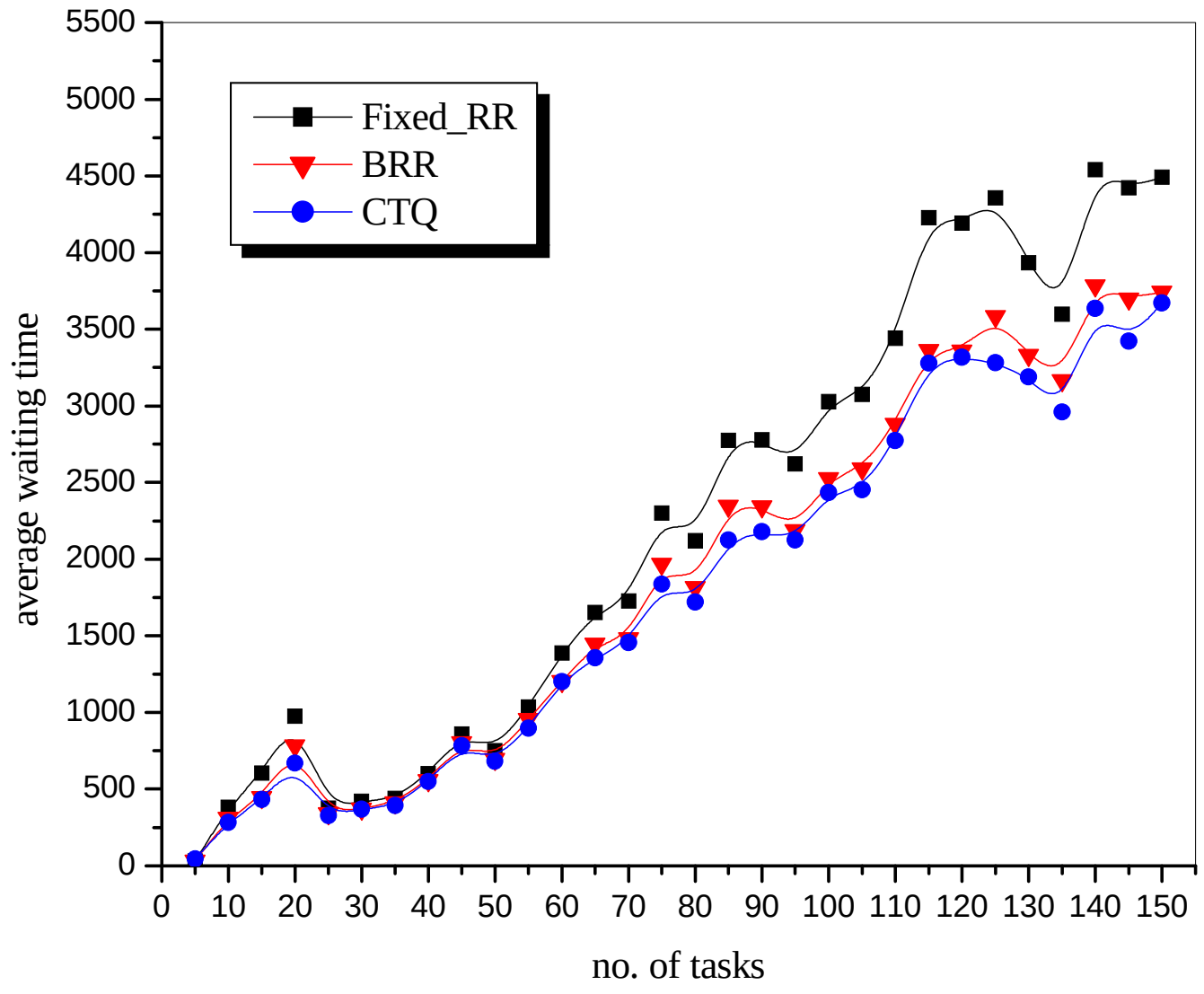

**Figures 4: Average waiting time for Fixed_RR, BRR and CTQ.**

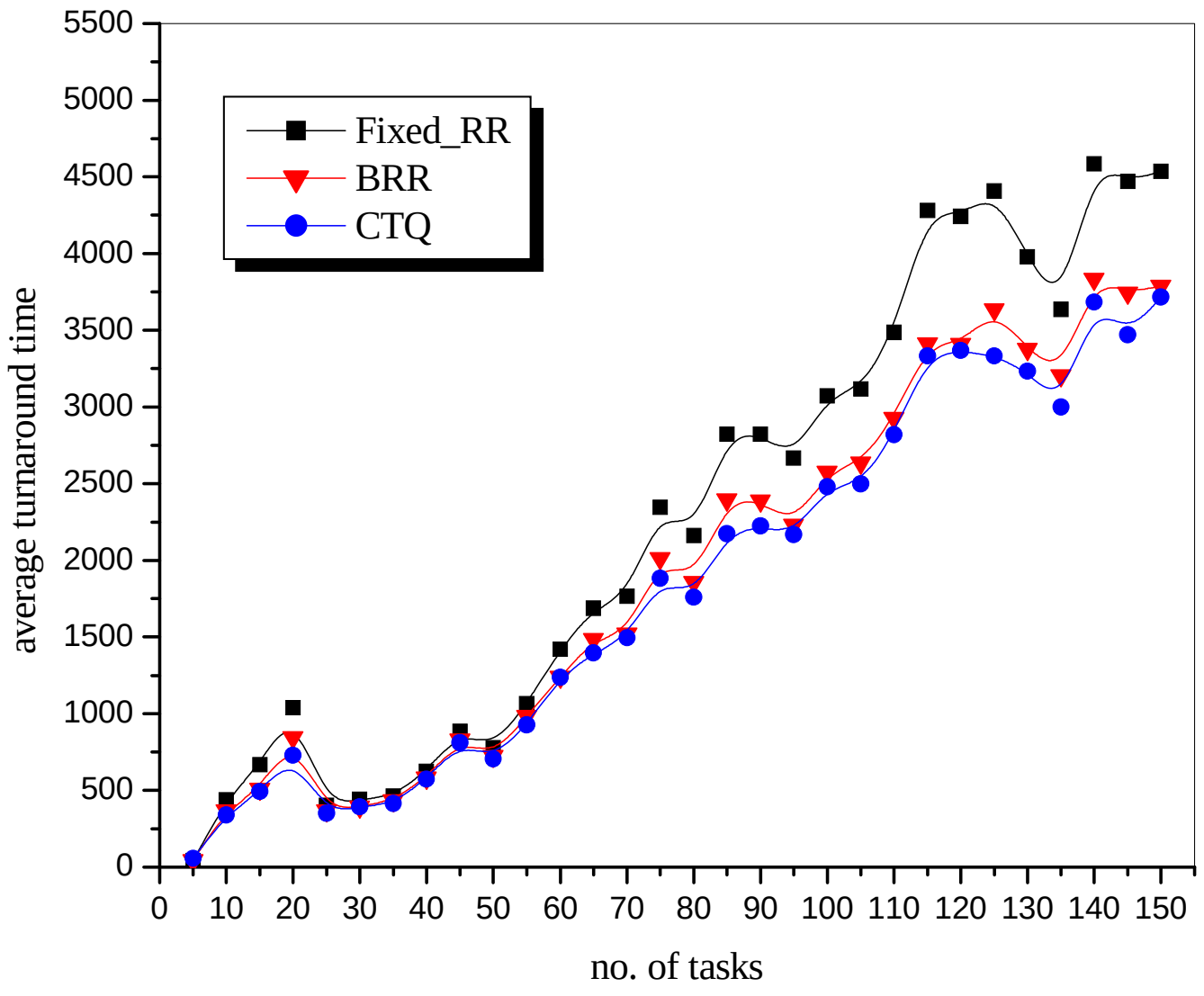

**Figures 5: Average turnaround time for Fixed_RR, BRR and CTQ.**

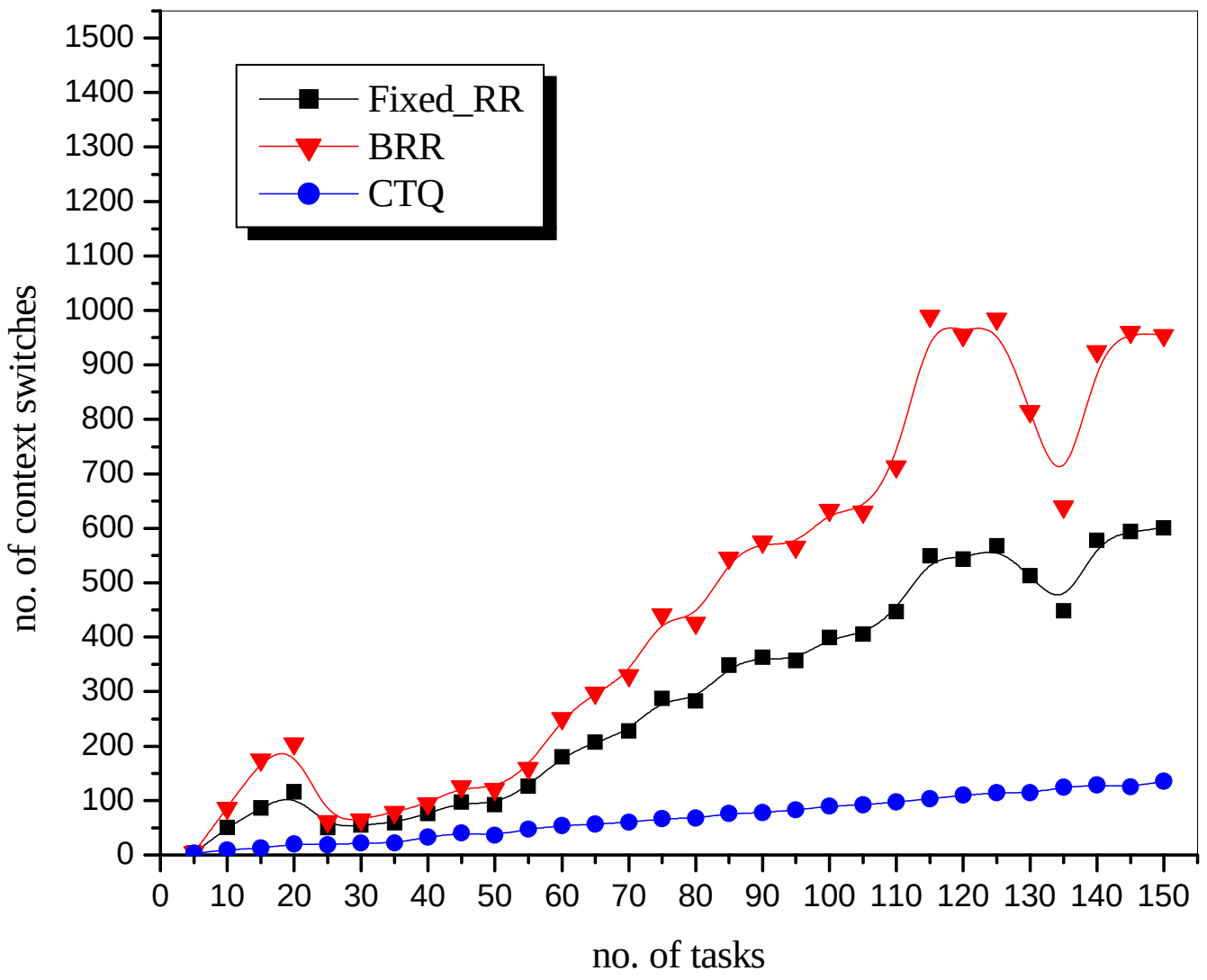

**Figures 6: Number of context switches for Fixed_RR, BRR and CTQ.**

## 1. CONCLUSION AND FUTURE WORK

In this view the round-robin has been studied as a function of the time quantum values. In this paper, we have proposed a new technique for round-robin scheduling algorithm based on process CPU burst time. In each round the algorithm adjusts the time quantum with respect to the residual times of the tasks and gives all possible $TQs$, each of which gives the smallest average waiting time. This technique has shown a good improvement over using a fixed time quantum RR**.** We introduced formulas which could be used to choose an appropriate time quantum value for practical use. As a future work, we would like to apply this technique with all possible round-robin variants. Another future work is to study the influence of selecting a time quantum in a round in the cyclic queue on the successive rounds and determine the relation between the current $TQ$ and the successive $TQs$.

[1] $\lfloor x \rfloor$ denotes the largest integer smaller than or equal to $x$.